\documentclass{agujournal2019}
\usepackage{array,tabularx,booktabs}
\usepackage{setspace}
\journalname{Perspectives of Earth and Space Scientists}

\begin{document}

\title{Funding Landscape of Early-Career Scientists: A Perspective from the AGU SPA Early Career Leadership Advisory Committee}

\authors{%
    Savvas Raptis\affil{1},
    Claire Gasque\affil{2},
    Samuel Hart\affil{3},
    Yi Qi\affil{4},
    Chris R. Gilly\affil{5},
    B. L. Alterman\affil{6},
    Ryan M. McGranaghan\affil{7},
    Juan Camilo Buitrago-Casas\affil{2},
    Komal Kumari\affil{8},
    Garima Malhotra\affil{9},
    Parisa Mostafavi\affil{1},
    Eduardo Perez Macho\affil{10},
    Onyinye Nwankwo-Omoloja\affil{11}
}

\affiliation{1}{Johns Hopkins University Applied Physics Laboratory, Laurel, MD, USA}
\affiliation{2}{Space Sciences Laboratory, University of California, Berkeley, CA, USA}
\affiliation{3}{Southwest Research Institute, San Antonio, TX, USA}
\affiliation{4}{Laboratory for Atmospheric and Space Physics, University of Colorado Boulder, Boulder, CO, USA}
\affiliation{5}{NorthWest Research Associates, Boulder, CO, USA}
\affiliation{6}{Independent Scientist}
\affiliation{7}{Jet Propulsion Laboratory, California Institute of Technology, Pasadena, CA, USA}
\affiliation{8}{Department of Electrical and Computer Engineering, University of Illinois Urbana-Champaign, Urbana, IL, USA}
\affiliation{9}{Cooperative Institute for Research in Environmental Sciences, University of Colorado Boulder, Boulder, CO, USA}
\affiliation{10}{Instituto Nacional de Pesquisas Espaciais, S\~{a}o Jos\'{e} dos Campos, Brazil}
\affiliation{11}{University of Michigan, Ann Arbor, MI, USA}

\correspondingauthor{Savvas Raptis}{savvas.raptis@jhuapl.edu}

\begin{keypoints}
    \item U.S. early-career funding in heliophysics is fragmented by career stage, nationality, and irregular solicitation schedules.
    \item Mid- and senior-career respondents point to predictable opportunity, not ease, as what enabled their independence.
    \item Predictable early-career PI programs and transparent selection data would strengthen career pathways for new scientists.
\end{keypoints}

\begin{abstract}
Like researchers across many scientific fields, early-career researchers in heliophysics face a critical transition from supervised research to independent scientific inquiry. However, the funding landscape that should support this transition is fragmented and unpredictable. Here we examine the current funding opportunities available to early-career researchers in heliophysics alongside the historical experiences of established scientists who navigated similar transitions. We catalog the major U.S. federal programs targeting early-career space physicists, documenting their eligibility constraints, selection rates, and accessibility. Access varies with career stage, appointment type, citizenship requirements, and solicitation schedules, and selection rates can swing sharply: the NASA Heliophysics Early Career Investigator Program (ECIP) selected 28\% of proposals in 2022 but only 6\% in 2024. To provide historical context, we collected survey responses from 30 researchers whose PhDs span 1972--2016 and conducted eight interviews with mid- and senior-career researchers. Most respondents viewed the present funding environment as more difficult than their own early careers, identifying predictability of opportunity, not ease, as what enabled earlier generations to build independence. Salary support alone does not guarantee opportunities to develop scientific leadership; accessible PI awards and real responsibility within collaborative projects can both contribute to that development. We recommend predictable, adequately sized grant programs for early-career principal investigators, transparent reporting of selection statistics across all early-career programs, continued use of dual-anonymous peer review, and named leadership roles for early-career researchers within large projects.
\end{abstract}

\section*{Plain Language Summary}
Early-career researchers in heliophysics, the study of the Sun and its effects on space and planetary environments, must make the transition from supervised research to leading their own. However, the funding that should support this transition is fragmented and unpredictable. We review the main U.S. federal programs for early-career space physicists, including who can apply and how many proposals are funded. Access depends on career stage, type of position, citizenship, and how often programs are offered, and selection rates vary widely. For example, NASA's Heliophysics Early Career Investigator Program funded 28\% of proposals in 2022 but only 6\% in 2024. We also surveyed 30 established researchers with PhDs from 1972 to 2016 and interviewed eight of them. Most viewed today's funding environment as more difficult than in their own early careers, and said that what helped them become independent was predictable opportunity rather than easy funding. We recommend regular, adequately funded grants for early-career researchers, public reporting of how many proposals are submitted and selected, continued anonymous review of proposals, and named leadership roles for early-career researchers within large projects.

\section{Introduction: The Early-Career Landscape}

A new space physics PhD has the technical and analytical skills to pursue scientific questions but not yet the means to pursue them independently. The ``early career'' stage, typically the first decade after graduation, is the transition from working within a supervisor's project to building an independent research portfolio. By its end, the researcher has developed new ideas, built collaborations,mentored students, and secured a steady pipeline of funding, but the path there is rarely clear.

Early-career (EC) researchers carry the field forward, becoming the scientific leaders and principal investigators who develop mission concepts, lead instrument development, and drive the resulting investigations. The experience they gain at this stage shapes the researchers who will later mentor others and maintain long-term expertise within the community. Mission concepts such as Interstellar Probe \cite{nas2024heliophysics} also envision lifetimes spanning several generations of researchers, which requires sustained expertise and long-term commitment from young leaders. Attrition at this career stage therefore costs more than individual careers. The field loses accumulated technical and institutional knowledge and a share of its future leaders and mentors. Studies of science funding support this concern. \citeA{bol2018matthew} showed that among recent PhDs with near-identical proposal review scores, those who narrowly won an early-career grant accumulated more than twice as much subsequent funding over the following eight years as those who narrowly missed the threshold, while also achieving higher rates of midcareer awards and full professorships. \citeA{wang2019early} found a complementary pattern among National Institutes of Health (NIH) applicants: a single early-career near miss predicted a more than 10\% chance of leaving the NIH funding system permanently, even though those who stayed later outperformed narrow winners. Together, these studies suggest that missing an early-career opportunity can permanently alter a career, through cumulative advantage for narrow winners, the ``Matthew effect'' \cite{bol2018matthew}, or through attrition of narrow losers \cite{wang2019early}. The transition to independence also often coincides with major personal decisions about where to live, whether to put down roots at an institution or in a region, and how to balance a career with family and other commitments. A career that repeatedly requires researchers to absorb funding gaps, relocate, or accept uncertain employment can become untenable even for scientists who strongly wish to stay in the field.

This perspective focuses on U.S. federal funding opportunities relevant to the Space Physics and Aeronomy (SPA) section of the American Geophysical Union (AGU), including opportunities open to eligible international researchers. Research within SPA encompasses solar physics, heliospheric and magnetospheric physics, and upper atmospheric and ionospheric physics at Earth and other bodies in our solar system. The AGU SPA Early Career Leadership Advisory Committee (EC-LAC), founded in 2022, is charged with understanding the experiences of early-career researchers within SPA and communicating them to SPA leadership. That work includes presenting statistics on the state of the SPA early-career community, compiling a list of early-career opportunities, and identifying gaps in those opportunities. It has also brought forward recurring concerns among early-career researchers, particularly about the stability and accessibility of research funding. SPA and the EC-LAC serve an international community, but the landscape analyzed here is mainly the U.S. system. Early-career researchers based elsewhere work within different national and regional funding structures, with their own eligibility rules, career-stage definitions, and challenges, which deserve separate treatment. We consider early-career researchers across postdoctoral, research-scientist, and faculty appointments.

In summer 2025, the EC-LAC conducted a community survey of early-career SPA researchers. The survey came at an uncertain moment for federal science funding. The FY26 President's Budget Request proposed deep reductions to federal science agencies, including a 47\% cut to NASA's Science Mission Directorate and a 46\% cut to Heliophysics, while NASA's ROSES-2025 research solicitation was released months after its traditional February date. The subsequent funding disruption further affected the ROSES-2025 schedule, with several proposal deadlines temporarily moved to ``TBD'' before being reset. In that survey, more than half of North American respondents reported that they wanted to remain in the field but feared that insufficient funding could force them to leave, while only 21\% expressed optimism about their long-term career prospects \cite{allen2026eclac}. Rapid-response surveys of U.S. early-career researchers in other fields report similar concern over career continuity and attrition following federal funding disruptions \cite{hammond2026early}.

Congress ultimately rejected most of the proposed FY26 cuts, providing some temporary stability to the federal science budget, but the uncertainty surrounding research funding did not disappear. The FY27 President's Budget Request again proposed substantial reductions to NASA Science, Heliophysics, and NSF relative to FY26 enacted levels, while NASA continued to operate on a shifted ROSES schedule. For researchers in soft-money positions (those whose salaries depend in whole or in part on external grants rather than institutional funds), grant success can decide whether their position continues. The period in which researchers are expected to establish scientific independence may therefore be the period in which they have the least buffer against disruptions in the funding system. Congressional appropriations can restore funding at the agency level, but they do not necessarily restore the predictable funding cadence on which individual research programs and early-career researchers depend.

The structure through which funding is distributed matters as much as the amount. As NASA increasingly supports large, collaborative investigations and changes how missions are supported beyond their prime phases, the balance of available opportunities may also change. In particular, fewer small grants, mission roles, or opportunities to lead individual investigations could make it harder for early-career researchers to gain the independence and leadership experience needed to establish their own research programs.

This Perspective has three goals. First, we give soon-to-graduate students and recent postdocs a non-exhaustive list of current funding opportunities that specifically target EC researchers, with selection rates where available. Second, we analyze a survey and a set of interviews with mid- and senior-career space physicists to describe the paths they took to becoming independent researchers. Third, using both the funding data and these accounts, we offer recommendations for keeping credible paths to scientific independence open.

\section{Current Funding Opportunities}

Table~\ref{tab:programs} summarizes the programs discussed below, their main eligibility constraints, published selection data, and sources.

\begin{table}[tbp]
\centering
\caption{Major U.S. funding opportunities relevant to early-career SPA researchers.}
\label{tab:programs}
\begingroup
\small
\setlength{\tabcolsep}{4pt}
\renewcommand{\arraystretch}{1.15}
\begin{tabularx}{\linewidth}{@{}
  >{\hsize=.80\hsize\linewidth=\hsize\raggedright\arraybackslash}X
  >{\hsize=1.25\hsize\linewidth=\hsize\raggedright\arraybackslash}X
  >{\hsize=1.05\hsize\linewidth=\hsize\raggedright\arraybackslash}X
  >{\hsize=.90\hsize\linewidth=\hsize\raggedright\arraybackslash}X@{}}
\toprule
\textbf{Program / agency} &
\textbf{Career stage / terms} &
\textbf{Citizenship / location} &
\textbf{Selection data / status} \\
\midrule
\textbf{ECIP}\newline NASA &
10-yr PhD window\textsuperscript{a}; no current or prior tenure\textsuperscript{b}; not offered annually &
No citizenship requirement; PI at U.S. institution &
ROSES-2024: 5 of 88 (6\%) \\
\addlinespace[4pt]
\textbf{YIP}\newline AFOSR &
7-yr PhD/equivalent window\textsuperscript{a} &
U.S. citizen, national, or PR; no DoD lab or FFRDC staff &
FY2024: 48 of 159 ($\sim$30\%); all disciplines \\
\addlinespace[4pt]
\textbf{NPP}\newline NASA/ORAU &
$<$5 yr from doctorate (otherwise Senior Fellow); initially 2 yr, up to 3 yr &
U.S. citizens, LPRs, and eligible foreign nationals (see text); resident research at NASA or affiliated host &
Selection rate not reported \\
\addlinespace[4pt]
\textbf{Jack Eddy Fellowship}\newline NASA LWS; UCAR/CPAESS &
PhD $\leq$3 yr preferred; 2 yr &
Open internationally; U.S. host required &
5 fellows each in 2023 and 2024; no call for 2025 or 2026 \\
\addlinespace[4pt]
\textbf{AGS-PRF}\newline NSF &
Normally $\leq$2 yr from PhD\textsuperscript{c}; up to 24 months &
U.S. citizen, national, or PR; eligible U.S. host &
Selection rate not reported \\
\addlinespace[4pt]
\textbf{CAREER}\newline NSF &
Untenured assistant professor or equivalent; $\geq$50\% tenure-track or equivalent appointment &
No citizenship requirement; eligible U.S. organization &
Not compiled here; several fields involved  \\
\addlinespace[4pt]
\textbf{Early Career Research Program}\newline DOE &
10-yr PhD window\textsuperscript{a}; untenured tenure-track asst./assoc.\ professor, or permanent, full-time, non-postdoctoral DOE lab/SC user-facility staff &
No citizenship requirement; eligible U.S. employer &
FY2026: 67 selectees; application total not reported \\
\addlinespace[4pt]
\textbf{GEO Core}\newline NSF 26-516 &
All career stages; proposals accepted anytime &
NSF organizational eligibility applies &
Too new to assess \\
\bottomrule
\end{tabularx}
\par\smallskip
\begin{minipage}{\linewidth}
\footnotesize\raggedright
\textit{Notes:}
Eligibility is only summarized here. Additional requirements and permitted exceptions are specified in the program rules. PR/LPR denotes lawful permanent resident; SC denotes DOE Office of Science. \textsuperscript{a}Degree windows follow competition-specific reference dates. The cited calls require degrees conferred on or after January 1, 2014 (ECIP, ROSES-2024), April 1, 2018 (YIP, FY2026), and January 1, 2015 (DOE, FY2026). DOE measures its window against the calendar year in which the federal fiscal year begins.
\textsuperscript{b}ECIP excludes current or previous holders of tenure or its non-academic equivalent, current or previous ECIP or PECASE recipients, and PIs/Science PIs of other current science or technology ROSES awards, except FINESST grants.
\textsuperscript{c}AGS-PRF also accepts graduate-student applicants and applicants more than two years from the PhD with less than 18 months of equivalent full-time employment at submission.
\par\smallskip
\textit{Sources:}
ECIP: \citeA{nasa_roses2024ecip,nasa_rosesstats2025};
YIP: \citeA{afosr_yip2023,afosr_yip2026};
NPP: \citeA{orau_npp2026};
Jack Eddy Fellowship: \citeA{cpaess_eddy2023,cpaess_eddy2024,ucar2026jackeddy};
AGS-PRF: \citeA{nsf_agsprf2022};
CAREER: \citeA{nsf_earlycareer2026};
DOE: \citeA{doe_ecrp2026,doe_ecrp2026awards};
GEO Core: \citeA{nsf_geocore2026}.
\end{minipage}
\endgroup
\end{table}

\subsection{NASA Early Career Investigator Program}

The NASA Heliophysics Early Career Investigator Program (ECIP) supports researchers within ten years of receiving their terminal degree. It is one of the few NASA programs specifically intended to help EC scientists establish independent research programs.

Recent competitions show how variable access can be. According to NASA's ROSES grant statistics \cite{nasa_rosesstats2025}, the 2024 Heliophysics ECIP competition received 88 Step-2 proposals and selected only five, a selection rate of approximately 6\%. In comparison, the 2022 competition selected 13 of 47 Step-2 proposals (28\%), the 2020 competition selected 14 of 54 proposals (26\%), and the 2018 competition selected 9 of 50 proposals for full funding (18\%), with three additional partial selections. Notably, the ROSES-2024 solicitation anticipated approximately 8--12 ECIP awards, whereas NASA selected only five. The published selection notice does not explain the shortfall.

\begin{figure}
    \centering
    \includegraphics[width=0.7\linewidth]{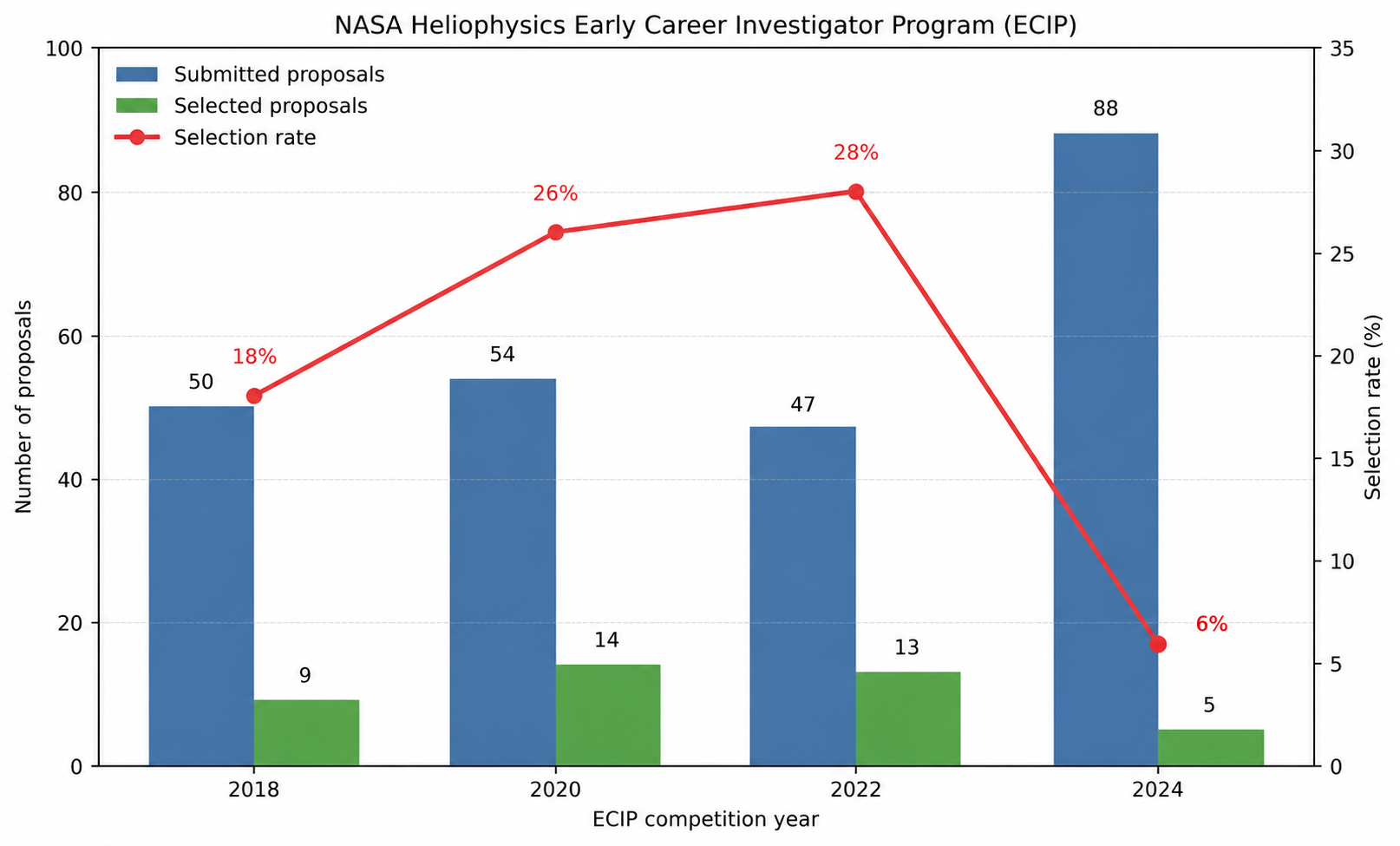}
    \caption{Recent NASA Heliophysics Early Career Investigator Program (ECIP) competitions. Bars show the numbers of submitted and selected proposals in each competition year, while the line shows the corresponding selection rate.}
    \label{fig1}
\end{figure}

ECIP's availability and selection rate have varied sharply between competitions, and the most recent competition fell to a single-digit selection rate (Fig.~\ref{fig1}). Since ECIP is not offered annually, an unsuccessful applicant may have to wait several years for another opportunity, during which they may approach or exceed the eligibility limit.

\subsection{Young Investigator Program}

The Air Force Office of Scientific Research Young Investigator Program (AFOSR YIP) provides three-year awards of up to \$450,000 for scientists and engineers within seven years of receiving their terminal degree. In FY2024, AFOSR received 159 proposals and announced 48 recipients, an overall selection rate of approximately 30\% across all supported disciplines. This aggregate rate does not show how many awards went to heliophysics-related research.

Access is also limited by citizenship, permanent residency, and employment requirements. The FY2026 solicitation specifies that the principal investigator must be a U.S. citizen, U.S. national, or permanent resident, must hold a full-time regular position at an eligible U.S. institution, and cannot be employed at a Department of Defense laboratory or federally funded research and development center. Applicants holding temporary visas are therefore ineligible, even when employed as tenure-track faculty or permanent research staff.

\subsection{NASA Postdoctoral Program}
The NASA Postdoctoral Program (NPP) offers research appointments at NASA centers and affiliated facilities. NPP eligibility includes U.S. citizens, lawful permanent residents, and foreign nationals eligible for J-1 exchange-visitor status, as well as specified applicants with pending immigration applications and valid employment authorization. Additional restrictions may apply to individual research opportunities. Applicants also need to identify a research opportunity, contact the associated NASA advisor, and develop a research proposal in consultation with that prospective advisor.

This structure can provide valuable mentoring and access to NASA expertise, but it also creates a substantial pre-application barrier. Success depends not only on the scientific proposal but also on identifying an appropriate opportunity and establishing an advisor relationship before submission. NPP accepts applications in multiple cycles each year, but recent program-wide application and selection totals are not publicly reported, so no reliable overall selection rate can be calculated.

Compared with many postdoctoral programs, NPP has less flexibility. Fellows are stipend recipients rather than employees of NASA, ORAU, or the host institution, and they are generally required to reside near and work at the approved NASA site. These conditions can create additional barriers related to relocation, benefits, family responsibilities, and immigration status.

\subsection{Jack Eddy Postdoctoral Fellowship}
The NASA Jack Eddy Postdoctoral Fellowship is a two-year fellowship sponsored by NASA's Living With a Star program and administered by UCAR/CPAESS. It supports interdisciplinary heliophysics research by matching recent PhD recipients with experienced researchers at U.S. host institutions. The program selected five fellows in both 2023 and 2024.

Although the fellowship is open to international applicants, fellows must identify a host scientist at a U.S. research institution while developing the application. The small number of appointments indicates limited capacity, but CPAESS does not publish application totals, so the selection rate cannot be independently assessed.

The program was paused for two years before returning in 2023 and has since paused again, with no positions offered for 2025 and no applications accepted for 2026 \cite{ucar2026jackeddy}. Such interruptions further reduce the already limited number of postdoctoral opportunities available to early-career researchers.

\subsection{NSF AGS Postdoctoral Research Fellowship}

The NSF Atmospheric and Geospace Sciences Postdoctoral Research Fellowship (AGS-PRF) supports up to 24 months of research in atmospheric and geospace sciences. Applicants must identify an eligible host institution and scientific mentor and develop both a research plan and a professional-development plan.

Although AGS-PRF offers greater choice of host institution than NPP, it is not fully flexible. Fellows are responsible for arranging their host affiliation and for covering expenses such as health insurance from the fellowship allowance, rather than automatically receiving standard employee benefits. The fellowship generally requires full-time research at the host institution and may therefore require relocation. Under NSF solicitation 22-639, applicants must be U.S. citizens, U.S. nationals, or permanent residents at submission, and may submit one proposal per year.

NSF reports individual awards but does not provide readily accessible annual application totals, so program-level selection rates cannot be consistently determined.

\subsection{Other Sources}

Additional programs become available only after researchers obtain particular appointments. NSF CAREER generally requires at least a 50\% tenure-track or tenure-track-equivalent assistant professor position, while the Department of Energy Office of Science Early Career Research Program is limited primarily to untenured tenure-track faculty and eligible DOE laboratory researchers.

During the writing of this Perspective article, NSF reorganized its core geospace research funding. In August 2026, NSF introduced Geosciences Core Research: Atmospheric, Geospace, Earth and Ocean Sciences (GEO Core; NSF 26-516), which consolidates and replaces the stand-alone GEM, SHINE, CEDAR, Geospace Cluster, and several other GEO funding opportunities \cite{nsf_geocore2026}. GEO Core accepts proposals at any time through its Atmospheric and Geospace Sciences program. Unlike programs reserved for early-career researchers, GEO Core is open across career stages and places no career-stage restriction on who may serve as PI, although NSF explicitly lists early-career investigators among the researchers it welcomes. Because this funding structure was introduced while this Perspective was being finalized, its implications for early-career access and program-specific selection rates cannot yet be evaluated.

\subsection{Early-Career Representation Among Recent Awards}
\label{sec:awards}
To examine how these funding opportunities translate into awards for early-career researchers, we compiled publicly available award records for NASA Heliophysics programs with selections announced in 2024--2025. Across the programs in our compilation, 939 proposals were submitted and 164 were selected, an overall selection rate of 17.5\%. We identified 60 selected PIs (37\% of all awardees) as early career, defined as having received their terminal degree within ten years of the selection year. Of these EC awardees, 15 (25\%) were within 8--10 years of receiving their terminal degree and therefore near the upper end of the EC eligibility window.




We performed a similar compilation of NSF awards made during 2024--2026 under the then-separate GEM, SHINE, and CEDAR solicitations. Early-career PIs accounted for 25 of 66 GEM awardees (38\%), 18 of 42 SHINE awardees (43\%), and 20 of 90 CEDAR awardees (22\%), including conference contracts and support. These categories are now historical, since GEO Core replaced the three stand-alone solicitations in 2026.

These numbers describe the composition of awardees, not EC proposal success rates. We do not know the EC fraction of submitted proposals in these recent competitions. NASA Heliophysics statistics from ROSES 2016--2021 show that EC PIs had selection rates similar to those of mid- and late-career PIs \cite{nasa_hpdyearbook2023}. Without corresponding applicant-stage data, these award fractions cannot establish whether EC investigators were overrepresented or underrepresented among successful proposals.

NASA has adopted dual-anonymous peer review (DAPR) in several programs in recent years, and DAPR may help reduce career-stage bias. In NASA's comparison of the 2019 pre-DAPR and 2020 DAPR Heliophysics Guest Investigators Open (HGIO) competitions, EC investigators increased from 34.4\% to 40.3\% of submissions and from 35.5\% to 46.7\% of awards \cite{nasa_dapr2026}. The EC share of awards exceeded their share of submissions in both cycles, with a larger difference in 2020. However, this comparison alone cannot establish whether DAPR caused the change. Comparable before-and-after statistics are not readily available for the NSF programs considered here.

\subsection{Outlook of Early-Career Opportunities}

Several valuable programs for EC scientists exist, but access is fragmented by career stage, appointment type, nationality, host or mentor requirements, and irregular solicitation schedules. Delays, hiatuses, and inconsistent announcement schedules disproportionately affect EC researchers, who often work under fixed-term appointments, short funding horizons, immigration constraints, and time-limited eligibility windows. The approximately 6\% selection rate for the 2024 Heliophysics ECIP competition shows how scarce these opportunities can become. In addition, the absence of consistent application and selection statistics for NPP, the Jack Eddy Fellowship, and AGS-PRF prevents the community from judging whether existing programs provide adequate routes to research independence.

\section{Testimonials from Mid- and Senior-Career Researchers}

To place the current early-career experience in historical context, in May 2025 we circulated a separate survey of mid- and senior-career researchers through several heliophysics community newsletters (e.g., SPA, SHINE, GEM, CEDAR). Thirty researchers, whose PhDs span 1972--2016, responded; 20 identified themselves and 10 responded anonymously. Participants quoted by name approved the wording and attribution of their quotations, and individuals named in the acknowledgments consented to being identified. Other responses are quoted without identifying details. We asked how they remembered the funding environment early in their careers, which opportunities helped establish their independence, and whether today's environment appears more or less difficult. We then conducted eight semi-structured interviews with mid- and senior-career researchers, using a set of questions about career transitions, funding, scientific independence, and changes in the field. Not every question was asked in every interview. The survey and interviews were self-selected and are not intended as a representative sample; rather, we use them to identify common career paths and points of historical contrast. The sample also reflects survivorship bias because it includes only researchers who remained in the field, and retrospective accounts may be affected by recall bias.

\subsection{1970s NASA History and Its Impact on Early-Career Researchers}

Some respondents noted that the post-Apollo era was a difficult time to begin a research career because of limited funding. Although NASA administrators James Webb and Thomas Paine sought to sustain the agency's Apollo-era momentum, the Vietnam War and competing domestic priorities placed pressure on NASA's budget. NASA's budget contracted substantially between its mid-1960s peak and the early 1970s \cite{levine1982managing,loc1972astronautics}. By September 1969, NASA's combined civil-service and contractor workforce had fallen from a peak of about 420,000 to about 190,000 \cite{stg1969postapollo}. Hiring restrictions reduced the number of new college graduates entering NASA while raising the average age of its workforce \cite{loc1972astronautics}. NASA programs also faced cancellations and delays \cite{levine1982managing}.

Despite this contraction, several landmark missions were developed and launched during the 1970s, including Helios, IMP 8, ISEE, and Voyager in the solar wind and magnetosphere, and Atmosphere Explorers C and E in the upper atmosphere. These missions shaped the field for decades. Their scientific success, however, did not guarantee stable research support for early-career researchers.

Several respondents described joining an established mission team or funded research group as an important route to research support. One scientist, who obtained a PhD in 1973, wrote, ``Connecting with one of the active research groups...with many opportunities for instrument development and data analysis...would put you on the track for funded research. Contributing actively to the lead investigator's project would put you on track for your own awards.'' Reflecting on their research group's funding, another scientist who obtained a space physics PhD in 1975 wrote, ``The group had grants from both NSF and NASA that funded most of the research. Some were associated with instrument investigations (IMP 8, Voyager, ISEE, AMPTE), some were theme associated. People did not write proposals to fund their specific research, but rather they contributed to the blanket proposals that funded the group.'' Other respondents described difficulty obtaining research support. One 1972 PhD wrote, ``Funding was almost nonexistent for myself and many colleagues...I needed to teach high school, then found a job as a lab technician before getting a low paid postdoc opportunity.'' One respondent also reflected on changes in research funding over their career, writing that the newer funding paradigm ``resulted in considerable inefficiency.''

\subsection{What Has Changed in the Path to Independence?}

The survey responses do not show that earlier generations simply had easier careers. Some interviewees recalled a severe shortage of faculty positions and uncertain soft-money employment, and some survey respondents identified advantages available today, including early-career-specific programs, improved career development, commercial-space opportunities, and broadly transferable computational skills. Nevertheless, most respondents viewed the present research-funding environment as more difficult. In our qualitative classification, 20 of 30 respondents characterized it as clearly more difficult, six gave mixed, similar, or conditional assessments, and four considered it slightly better or less difficult. What respondents emphasized most was predictability. Respondents appear to have tolerated individual proposal failures because they expected another opportunity to follow, whether through another solicitation, mission, facility, postdoctoral program, or source of institutional funding. Recalling the uncertainty of his own early career, Prof. Jacob Bortnik (PhD 2004) described an important difference: ``...it was scary, but it worked. The system worked.''

Respondents also identified a change in how opportunities for leadership are distributed. Large team awards can support early-career scientists and enable ambitious research, just as missions and large projects did for many of the researchers we interviewed. However, several respondents worried that fewer small PI-scale opportunities make it harder for early-career researchers to move from contributor to scientific leader. Large projects are more likely to be led by established investigators, with access to leadership roles depending in part on existing networks. As Prof. Bortnik described this risk, ``you become a lot more like a worker, like an employee than a scientific leader.''

\subsection{How Earlier Careers Built Scientific Independence}

Early-career researchers who became established in our field did not follow a single path. Respondents described difficult job markets, low proposal success rates, soft-money uncertainty, and periods of sparse funding. What most of them had, however, was a runway long enough to develop as a scientist: missions, observatories, group-level grants, and postdoctoral fellowships often provided several years of continuous support while researchers built expertise, networks, and proposal experience.

Scientific independence often followed when that support was converted into a clear leadership role. Survey respondents repeatedly identified a first PI-scale grant or fellowship as a major career transition, including NASA postdoctoral programs and ECIP, along with NSF awards through programs such as SHINE and CAREER. Interviewees similarly described independent funding as the point at which they could pursue their own questions, lead collaborators, and show that their ideas could sustain a full research program. Dr. Sophie Musset (PhD 2016) described her first successful NASA proposal as the moment when she felt she had ``proved that my ideas are viable and can be funded,'' and could demonstrate that she was leading her own project. Mentors and senior colleagues often helped bridge these stages by making introductions, giving visible responsibility, and bringing junior researchers into missions and proposals.

Salary support alone does not provide the experience needed to become a future PI, group leader, or mission scientist. Respondents emphasized that early-career researchers also need time to develop their own ideas, senior scientists willing to hand over responsibility, and opportunities to demonstrate leadership. Several respondents suggested giving early-career researchers named responsibility for scientific tasks or work packages within large collaborations.

\section{Discussion and Recommendations}

The summary of funding programs documented in Section~2 and the career histories described in Section~3 point to the same conclusion: the current system can support individual early-career researchers, but taken together it does not provide a reliable path from postdoc to independent scientist. Earlier generations faced difficult job markets and low success rates, yet most respondents recalled confidence that a viable next opportunity would exist. Today, irregular solicitation schedules, biennial or interrupted programs, undisclosed programmatic reasoning in selections, and sharp swings in selection rates (ECIP, from 28\% in 2022 to 6\% in 2024) make it difficult for early-career researchers to plan a multi-year strategy for establishing independence, precisely when their fixed-term appointments and eligibility clocks leave the least room for delay. As of October 2026, when this article was being finalized, the situation had not improved. NASA had indicated that ROSES-2026 would be released in July 2026, but it never appeared \cite{nasa_roses2026}. In October 2026, NASA instead announced that ROSES-2027 would be released no later than January 2027, with ROSES-2025 remaining open until then \cite{nasa_roses2027}. Because NASA did not meet its own stated date for ROSES-2026, applicants cannot yet plan around the new one. Even if NASA meets it, 18 months will separate consecutive ROSES releases. For a researcher on a two-year postdoctoral appointment, that gap can consume most of the time available to develop, write, and submit a proposal before eligibility expires. The delay can also change career decisions outright. A researcher whose visa depends on continued employment, or who is a household's primary earner, may have to commit to a position outside research before learning whether a suitable opportunity will exist at all.

When independent funding is not available, early-career researchers typically continue working through other means: contributing to larger missions, joining established research groups, or accepting positions funded by a senior investigator's grant. These roles can provide both salary support and opportunities to develop scientific leadership, particularly when they carry authority over the research. Without that authority and time for independent ideas, continued employment may not lead to independence.

Short funding cycles, precarious employment, and hierarchical dependence may also push early-career researchers toward safer, more readily evaluated questions rather than higher-risk or more innovative work \cite{haubrock2026contemporary}. Predictable funding and scientific independence may therefore also support more innovative research. Dependence on a single senior PI for continued employment also creates a power imbalance that is rarely discussed. An early-career scientist whose position depends on one investigator's renewal has limited ability to advocate for lead authorship, pursue independent ideas, or raise concerns about working conditions.

These dynamics are not unique to heliophysics, but they are amplified in a field where independent funding opportunities are few and irregularly available. The testimonials in Section~3 reinforce this point: respondents who built successful careers consistently identified a first PI-scale grant or fellowship as the turning point, the moment when they could demonstrate that their ideas were independently viable.

Citizenship and residency rules narrow access further. AFOSR YIP and AGS-PRF exclude applicants on temporary visas, whereas NPP admits eligible foreign nationals subject to immigration and opportunity-specific requirements. Assessments of access for international researchers should treat these programs separately.

Drawing on the funding data and respondents' accounts, we make the following four recommendations.

\noindent\textbf{1. Provide predictable, adequately funded opportunities for early-career PIs.}
Dedicated grants for early-career principal investigators would keep independent work possible at the start of a career. Several respondents identified a first PI-scale grant as a key transition toward independence, yet ECIP is offered irregularly and reached a 6\% selection rate in its most recent cycle. This concern is consistent with the cumulative advantage described in Section~1 \cite{bol2018matthew}, part of which arose because near-miss applicants stopped applying for subsequent grants altogether. When a field offers few and irregular early-career competitions, the consequences of a single unsuccessful application can last well beyond the immediate funding cycle.

Accordingly, NASA and NSF should maintain predictable annual opportunities for early-career investigators to lead funded research, through strengthened existing programs or a dedicated mechanism. Awards should provide sufficient research time over multiple years to develop and lead a project, balancing award size and duration against the number of researchers supported. Awards that are too small or too short may require recipients to spend excessive time seeking additional funding. Agencies should publish anticipated solicitation and decision dates, keep application requirements proportionate to award size, and provide eligibility extensions when competitions are delayed or suspended. Evaluation should track first-time PIs supported, time to award, and subsequent opportunities for scientific leadership.

\noindent\textbf{2. Publish selection statistics and explain funding decisions.}
The absence of consistent application and selection statistics across programs limits the community's ability to assess early-career access. ECIP provides selection rates because NASA publishes ROSES grant statistics, but comparable numbers are not publicly available for NPP, the Jack Eddy Fellowship, or AGS-PRF. Programs targeting early-career researchers should publish annual application totals and selection rates. Programs open to multiple career stages should also report application and selection counts by career stage. When programmatic considerations influence selections, agencies should explain those considerations and distinguish them from the scientific assessment in applicant feedback. This would help applicants understand funding decisions and prepare future proposals.

\noindent\textbf{3. Continue and evaluate dual-anonymous peer review.}
We recommend continued use and evaluation of dual-anonymous peer review (DAPR), which is designed to reduce the influence of investigator identity on scientific assessment. The HGIO comparison in Section~\ref{sec:awards} is encouraging, but two competition cycles cannot separate an effect of DAPR from changes in applicant pools, proposal quality, funding, or other factors. Evaluation should therefore draw on application and selection counts by career stage over multiple cycles.

\noindent\textbf{4. Support early-career leadership within larger projects.}
Large-project funding should complement independent awards through named, funded leadership responsibilities for early-career researchers. These roles should provide authority to define research questions, manage resources, and lead collaborators, together with protected time to develop independent ideas.

Community efforts such as leading conference sessions, organizing working groups and focus groups, and taking on leadership roles within the GEM, SHINE, and CEDAR communities and AGU committees can build leadership skills and professional visibility. That experience is valuable, but it is typically unpaid and does not substitute for running an independent, funded research program, which is needed to advance from an early-career stage to a successful mid-career one. Formal recognition of such service in proposal evaluation criteria would help bridge the gap between community leadership and research independence.

Finally, the consequences of an unstable career path extend beyond individual researchers. Some heliophysics capabilities depend on technical and scientific expertise accumulated over many years, and breaks in career continuity can become losses of expertise that are difficult to rebuild. As Dr. Philip Erickson (PhD 1998) argued, early-career scientists ultimately ``have to have something to shoot for, which is some kind of permanence.''

\section*{Open Research Statement}
The survey and interviews described in this manuscript collected professional perspectives from mid- and senior-career researchers. All interviewees gave written or verbal consent to the interpretation and publication of their views. Participants quoted by name approved the wording and attribution of their quotations before submission, individuals named in the acknowledgments consented to being identified, and all others remain anonymous. The NASA and NSF award records used in Section~\ref{sec:awards} are publicly available from the sources cited in the text.

\section*{Conflict of Interest Disclosure}

The authors declare there are no conflicts of interest for this manuscript.

\acknowledgments
We thank the 30 members of the heliophysics community who responded to our survey, including Rob Ebert, K. D. Leka, Sophie Musset, Don Mitchell, Angelos Vourlidas, Jacob Bortnik, John C. Foster, Matthew Fillingim, Christina Cohen, Allison Jaynes, Angeline Burrell, Preeti Bhaneja, Katelynn Greer, Hyunju Connor, Kathy Reeves, Mark Moldwin, Philip Erickson, and Amy Keesee, as well as those who remained anonymous. We also thank Philip Erickson, Jacob Bortnik, John Foster, K. D. Leka, Mark Moldwin, Shea Hess Webber, Angelos Vourlidas, and Sophie Musset for sharing their experiences through follow-up interviews. We also thank AGU, and specifically the SPA section leadership, for their continued support of early-career scientists.

\bibliography{wiley}

\end{document}